\documentclass{PoS}

\title{Pion-pion interaction in the I = 1 channel }

\ShortTitle{Pion-pion interaction in the I = 1 channel }

\author{\speaker{Bruno Charron} \\
        Physics Dept., University of Tokyo, 7-3-1 Hongo, Bunkyo, Tokyo 113-0033, Japan\\
        E-mail: \email{charron@riken.jp}}

\author{for HAL QCD Collaboration}

\abstract{We present preliminary results of a new approach to the study of the pion-pion system in the I=1 channel. The Bethe-Salpeter wave function of the two-pion system is computed on the ground state and the first excited state. From these, we attempt to extract an interaction kernel (potential) which can then be used to extract observables such as the phase shifts. In a first trial, we use rather large pion masses $m_\pi \sim 1.05$ GeV and $m_\pi \sim 0.68$ GeV which do not allow rho decay.}

\FullConference{The 30th International Symposium on Lattice Field Theory\\
                 June 24 - 29,  2012\\
                 Cairns, Australia}

\usepackage{amsmath}
\usepackage{amssymb}

\usepackage{caption}
\usepackage{subcaption}

\begin{document}

\section{Introduction}

In recent years, the I=1 two-pion system has attracted a lot of attention in lattice QCD. The increase in computational power has finally allowed to generate fully dynamical QCD gauge configurations at quark masses low enough for the rho resonance to be observed, promising a better understanding of complex hadron processes from first principles. Recent studies \cite{pacscs-rho, lang-rho, bmw-rho, feng-rho} all use the de facto standard method, which is to apply L\"uscher's formula \cite{luescher} or its extension to moving frames \cite{rummukainen}, in order to relate the finite-size energy spectrum to the infinite volume phase shifts. The main difficulty faced by this method is that one can only extract the phase shifts at a few energies on the lattice, making it difficult to reconstruct the continuous energy range and thus the physical parameters of the system, especially around the resonance.

An alternative method has been recently introduced for the study of the nucleon-nucleon system \cite{hal-proposal} then successfully extended, in particular, to various baryon-baryon systems \cite{hal-others}. The method relies on the fact that the phase shifts can be obtained from the asymptotic behaviour of the Bethe-Salpeter (BS) wave functions. An effective energy-independent, non-local potential can then be introduced to account for the energy dependence of the BS wave functions, and as a result, the energy dependence of the phase shifts.

This paper reports on our first attempt to apply this potential method to the I=1 two-pion system. The meson masses considered here do not allow for the rho meson to decay, the goal being to test the viability of the method in this channel before applying it to the study of the resonance.

\section{BS wave function and potential}

To describe two pions in the isospin $I=1$ channel, we use the following operator
\begin{align}
  \label{def_pipi_op}
  \pi\pi(\textbf p) &= \frac{1}{\sqrt{2}} \left[ \pi^- (\textbf p) \pi^+ (-\textbf p) - \pi^+ (\textbf p) \pi^- (-\textbf p) \right],
\end{align}
where $\pi^\pm$ are local interpolating operators for the pions. The Bethe-Salpeter (BS) wave function in the center of mass frame is then defined in this channel as
\begin{align}
  \label{def_bs_wf}
  \Psi_n(\textbf r) = \int \frac{d^3\textbf p}{(2\pi)^3} e^{i \textbf r \cdot \textbf p} \langle 0 \vert \pi\pi(\textbf p) \vert n \rangle
\end{align}
with $\vert n \rangle$ an eigenstate of QCD with the required quantum numbers. The BS wave function presents an asymptotic behaviour \cite{hal-theory} which allows to extract the scattering phase shift at the energy of the eigenstate, in exactly the same way as the wave function in quantum mechanics.

By inversion of the energy-dependence of the wave functions \cite{hal-theory}, one can define a non-local energy-independent potential $U$ such that the wave functions $\Psi_n$ satisfy, for all eigenstates $\vert n \rangle$ with energies below the inelastic threshold, the Schroedinger-like equation
\begin{align}
  \label{eq:schrod_eq}
  (\boldsymbol \nabla^2 + \textbf k_n^2) \Psi_n (\textbf r) = m_\pi \int d^3\textbf r' \; U(\textbf r, \textbf r') \Psi_n (\textbf r').
\end{align}
where the energy of the eigenstate $\vert n \rangle$ is $E_n = 2\sqrt{\textbf k_n^2 + m_\pi^2}$.

Since we only have access to a limited number of such BS wave functions on the lattice, we can only obtain an approximation of the potential, such as a truncation of its derivative expansion. The velocity expansion of the non-local potential in this channel is
\begin{align}
  \label{eq:der_exp}
  U(\textbf r, \textbf r') = [V_0(\textbf r) + V_2(\textbf r) \boldsymbol \nabla^2_{\textbf r'} + \dots ] \delta(\textbf r' - \textbf r).
\end{align}
From the approximate potential, we can obtain approximate BS wave functions at all energies below the inelastic threshold and extract phase shifts from them. The accuracy of the results thus depend on the convergence of the derivative expansion.

\section{Wave functions on the lattice}

\subsection{Eigenstates and correlation matrix \label{sec:eigenstates}}

We note $\mathcal{O}(t)$ a functional of the quark fields at time $t$, and $\mathcal{O}$ its corresponding operator in the Heisenberg picture. In the limit of an infinite lattice in the time direction, we have
\begin{align}
  \label{eq:corr_func}
  \langle \mathcal{O}_1(t) \mathcal{O}_2(t_0)\rangle = \sum_n \langle 0 | \mathcal{O}_1 | n\rangle \langle n | \mathcal{O}_2 | 0\rangle e^{-E_n(t-t_0)}.
\end{align}
where the brackets in the left-hand side denote the expectation value in lattice QCD.

Let $\{\mathcal{O}_i\}$ be a set of $N$ linearly-independent operators and $M_{nj} = \langle n | \overline{\mathcal{O}_j} | 0 \rangle$ the $N\times N$ matrix of their mixing with the $N$ lowest eigenstates. Their correlation matrix $G$ is defined as
\begin{align}
  \label{eq:corr_mat}
  G_{ij}(t, t_0) = \langle \mathcal{O}_i(t) \overline{\mathcal{O}_j}(t_0)\rangle = M^\dagger D(t - t_0) M + O(e^{-E_{N+1}(t-t_0)})
\end{align}
where $D$ is diagonal with components $D_{nn}(t-t_0) = e^{-E_n(t-t_0)}$.

Diagonalizing $G^{-1}(t', t_0)G(t, t_0)$ for several $(t, t')$ pairs yields the energies, by fitting of the eigenvalues with the form $e^{E_n(t'-t)}$, and the inverse mixing $M^{-1}$ as the eigenvector matrix (up to a normalization of the columns). This requires $t$ and $t'$ to be sufficiently separated from $t_0$ so that contributions from eigenstates higher than $|N\rangle$ vanish in \eqref{eq:corr_mat}.

The determination of $M^{-1}$ and the energies allows to compute, for any operator $\mathcal{O}$ and eigenstate $n\leq N$, the matrix element
\begin{align}
  \label{eq:combination}
  \langle 0 | \mathcal{O} | n\rangle = e^{E_n(t - t_0)} \sum_{i \leq N} (M^{-1})_{i,n} \langle \mathcal{O}(t) \overline{\mathcal{O}_i}(t_0)\rangle
\end{align}
and thus the BS wave functions as seen in the definition \eqref{def_bs_wf}.

\subsection{Source operators}

The pion-pion I=1 channel contains both pion-pion scattering states and the rho meson. In the pion mass region we investigate ($m_\pi$ = 1.05 and 0.68 GeV), we expect the ground state to be the rho meson, the first excited state to be the pion-pion scattering state with the lowest non-zero momentum allowed on the lattice, and other eigenstates to have energies large enough to only consider the first two ($N=2$).

To approximate the pion-pion state we use the operator $\mathcal{O}_1 = \pi\pi(\textbf p)$ with momentum $\textbf p = \frac{2\pi}{L} \textbf e_z$ ($L$ being the spatial extent of the lattice) and for the rho meson the operator
\begin{align}
  \mathcal{O}_2 = \rho = \frac{1}{\sqrt{2}} \sum_{\textbf x} \left[ \bar u (\textbf x) \textbf a \cdot \boldsymbol \gamma u (\textbf x) - \bar d (\textbf x) \textbf a \cdot \boldsymbol \gamma d (\textbf x) \right]
\end{align}
with a polarization taken parallel to that of the relative momentum of the pions, $\textbf a = \textbf e_3$.

\subsection{Correlation functions}

\begin{figure}
  \centering
  \def\svgwidth{.7\columnwidth}
  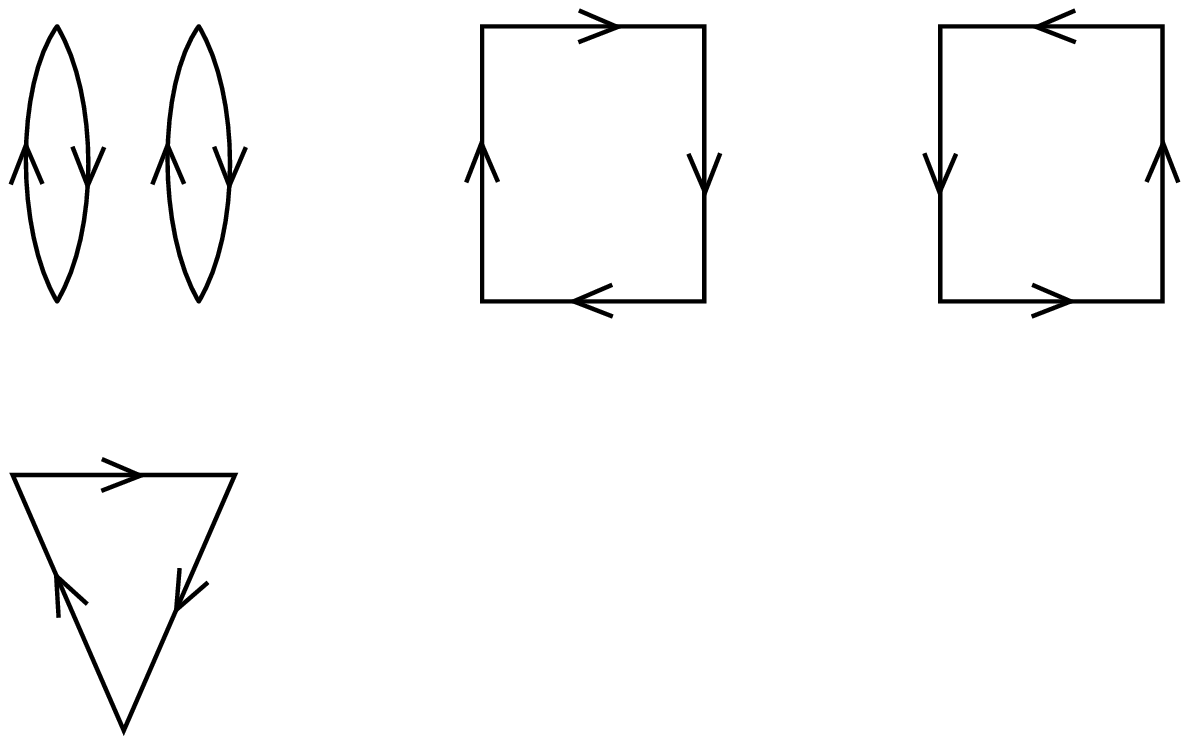
  \caption{Decomposition in Wick contractions of the correlation functions corresponding to $\pi\pi\rightarrow\pi\pi$ and $\rho\rightarrow\pi\pi$, appearing both in the correlation matrix (with $\textbf q = \textbf p$) and the wave function. Time goes upward.}
  \label{fig:diagrams}
\end{figure}

The correlation matrix and the wave functions require the evaluation of correlation functions of the form of \eqref{eq:corr_func}, some of which are illustrated in Fig.~\ref{fig:diagrams}. For the wave functions, the substitution $\textbf q \leftrightarrow -\textbf q$ translates to parity conjugation, so we only need to compute the first parts. The diagrams are computed, following \cite{pacscs-rho}, using adequate contractions of direct and sequential propagators using stochastic noises $\xi_j$ as sources
\begin{align}
  Q(\textbf x, t | \textbf q, t_S, \xi_j) &= \sum_{\textbf y} D^{-1} (\textbf x, t ; \textbf y, t_S) [e^{i\textbf q \cdot \textbf y} \xi_j(\textbf y)] \\
  W(\textbf x, t | \textbf k, t_1 | \textbf q, t_S, \xi_j) &= \sum_{\textbf z} D^{-1} (\textbf x, t ; \textbf z, t_1) [e^{i\textbf k \cdot \textbf z} Q(\textbf z, t_1 | \textbf q, t_S, \xi_j)].
\end{align}

\begin{figure}
        \centering
        \begin{subfigure}[b]{0.3\textwidth}
                \centering
                \includegraphics[height=.09\textheight]{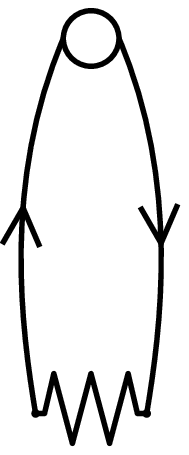}
                \caption{}
                \label{fig:contraction1}
        \end{subfigure}
        \begin{subfigure}[b]{0.3\textwidth}
                \centering
                \includegraphics[height=.09\textheight]{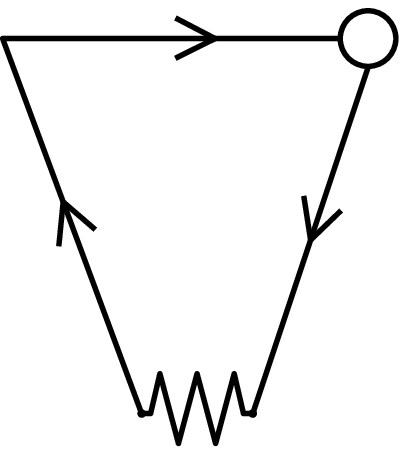}
                \caption{}
                \label{fig:contraction2}
        \end{subfigure}
        \begin{subfigure}[b]{0.3\textwidth}
                \centering
                \includegraphics[height=.09\textheight]{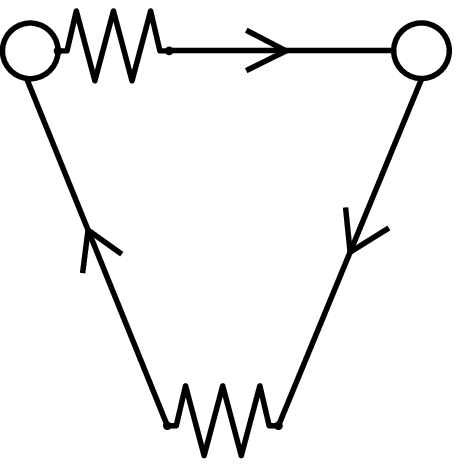}
                \caption{}
                \label{fig:contraction3}
        \end{subfigure}
        \caption{Computation method for some contractions. Springs link pairs of points which are projected one on the other by summing over stochastic noises. Open circles are explicit summations. Straight (resp. broken) arrows are direct (sequential) propagators.}
        \label{fig:contractions}
\end{figure}
Fig.~\ref{fig:contraction1} shows an example. The propagators are contracted explicitely at the sink (upper part in the diagram) and implicitely at the source (lower part) by noise projection. In Fig.~\ref{fig:diagrams} appear rectangle- and triangle-like diagrams. They can be computed in the same way as the previous example using sequential propagators, cf. Fig.~\ref{fig:contraction2}. However, while the momentum can be introduced freely at the explicit summation (empty circle), sequential propagators have definite intermediate momenta. This means that for the wave functions, which requires all possible sink momenta $\textbf q$, we need to compute as many sequential propagators. To remedy this, we introduce for the wave functions another stochastic noise at the sink, cf. Fig.~\ref{fig:contraction3}, which allows to choose the two momenta at the sink independently of the computation of the propagators. 

\section{Numerical setup}

The preliminary results presented here were computed using the $N_f=2+1$ full QCD gauge configurations of ILDG/JLDG generated by the CP-PACS and JLQCD Collaborations \cite{configs} on a $28^3\times 56$ lattice with a RG improved gauge action at $\beta = 2.05$ and a $O(a)$ improved Wilson quark action with $c_{SW} = 1.628$. The lattice spacing is $a = 0.0685$ fm which makes for a lowest non-zero momentum of $p = 2\pi / L = 0.65$ GeV. We compare our results on two sets of configurations with light quark hopping parameters $\kappa_{ud} = 0.1347$ and $\kappa_{ud} = 0.1356$, keeping $\kappa_{s} = 0.1351$ fixed.

The configurations with $\kappa_{ud} = 0.1347$ have meson masses $m_\pi = 1.05$ GeV and $m_\rho = 1.37$ GeV. Those with $\kappa_{ud} = 0.1356$ have masses $m_\pi = 0.68$ GeV and $m_\rho = 1.10$ GeV. In both cases, the lowest energy of two free pions in the center of mass frame, $2\sqrt{m_\pi + (2\pi/L)^2}$, is significantly larger than that of the rho meson at rest.

The quark propagators are computed with temporal Dirichlet boundary condition. We use $U(1)$ stochastic noises, 6 at the source and 20 at the sink. Wave functions are projected in the $T_1^-$ representation of the cubic group. Statistical errors are computed using the jackknife technique although 2-dimensional plots are shown without error bars for clarity.

\section{Preliminary results}

\begin{figure}
    \centering
    \includegraphics[width=.49\textwidth]{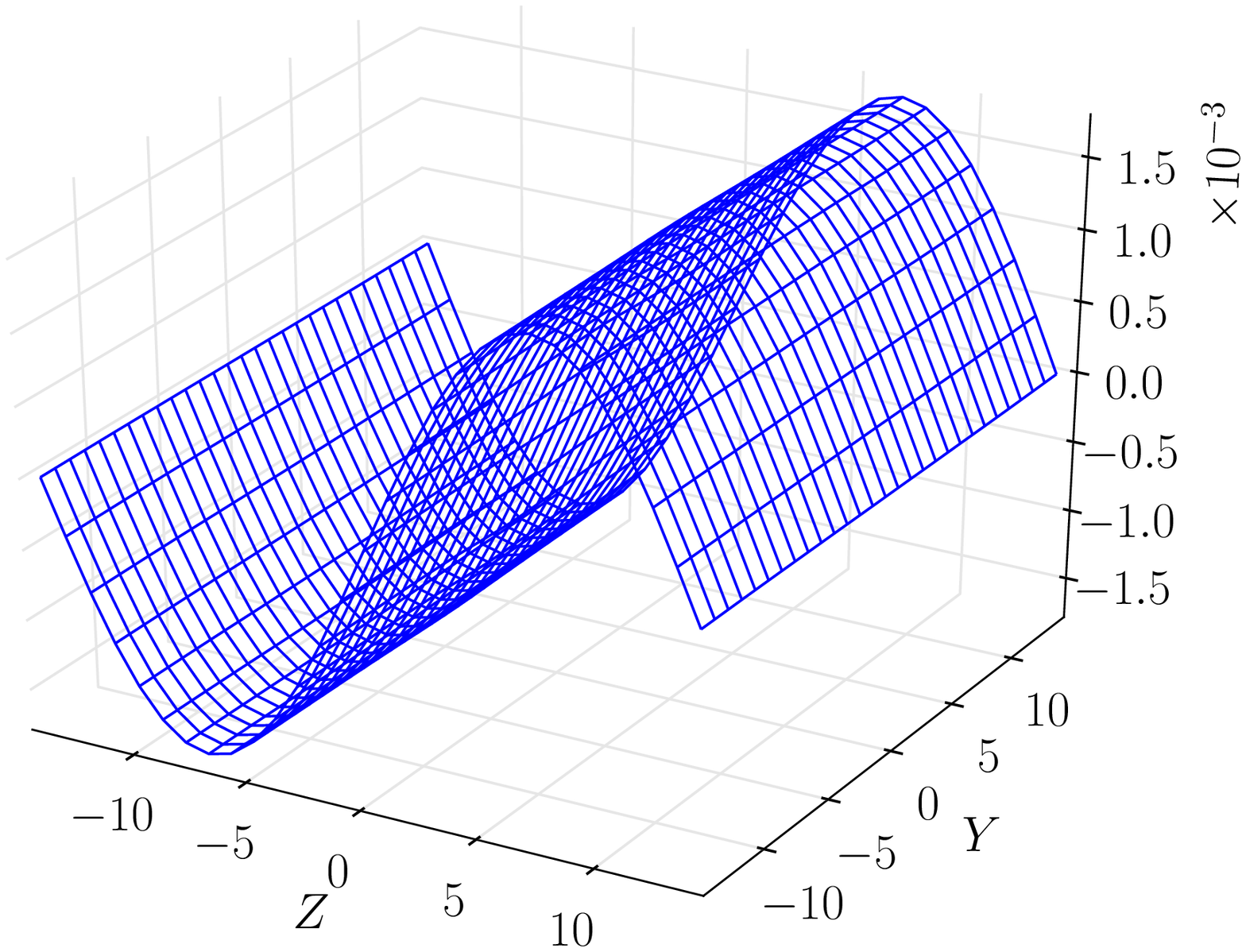}
    \includegraphics[width=.49\textwidth]{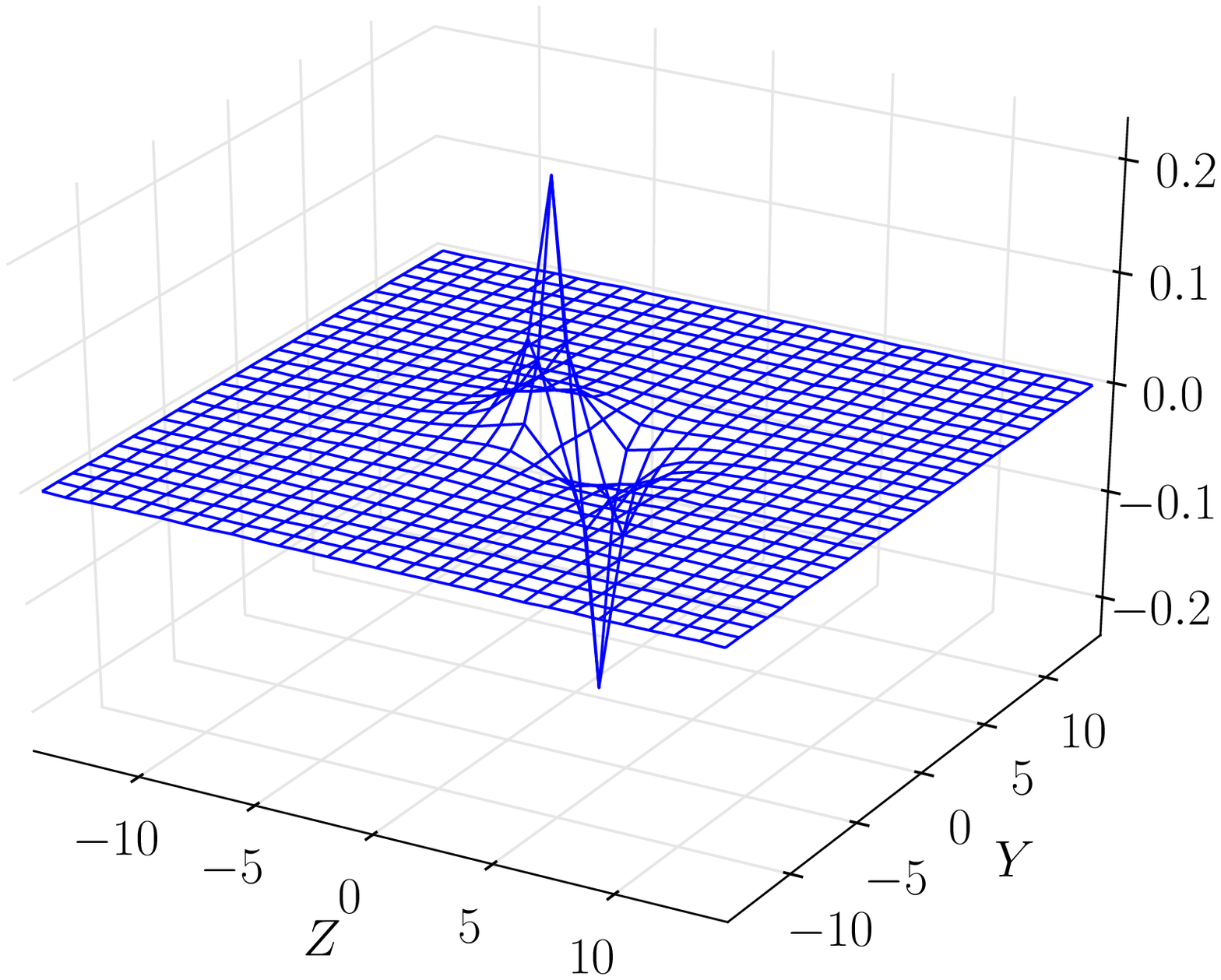}
    \caption{Wave functions for the first (left) and second (right) diagrams of the $\pi\pi\rightarrow\pi\pi$ correlation function in Fig.~\protect\ref{fig:diagrams} (upper). Normalized such that the total wave function has a norm $1$. Computed at $t-t_0 = 12$ on the set $\kappa_{ud} = 0.1347$.}
    \label{fig:diag_wf}
\end{figure}

We have seen that the wave functions \eqref{def_bs_wf} on the eigenstates are obtained as combinations \eqref{eq:combination} of the wave functions computed with the interpolating source operators $\pi\pi(\textbf p)$ and $\rho$, themselves computed as sum of Wick contractions ("diagrams"). The combinations are obtained by diagonalization of the correlation matrix.

Figure \ref{fig:diag_wf} shows the contribution of the two kind of diagrams appearing in the $\pi\pi\rightarrow\pi\pi$ wave function. The left one, corresponding to the "parallel" diagram, is close to the free wave function. The right one, corresponding to the rectangle diagram, exhibits a very peaked and short-ranged behaviour. The triangle diagram, Fig.~\protect\ref{fig:diagrams} (lower), has a wave function very similar (up to a normalization) to that of the rectangle one. The rho meson being the ground state, the quark-antiquark pair propagating in the rectangle and triangle diagrams from $t$ to $t_0$ can be thought as forming a rho meson, which could explain why the two diagrams' wave functions are similar and short-ranged.

\begin{figure}
    \centering
    \includegraphics[width=.49\textwidth]{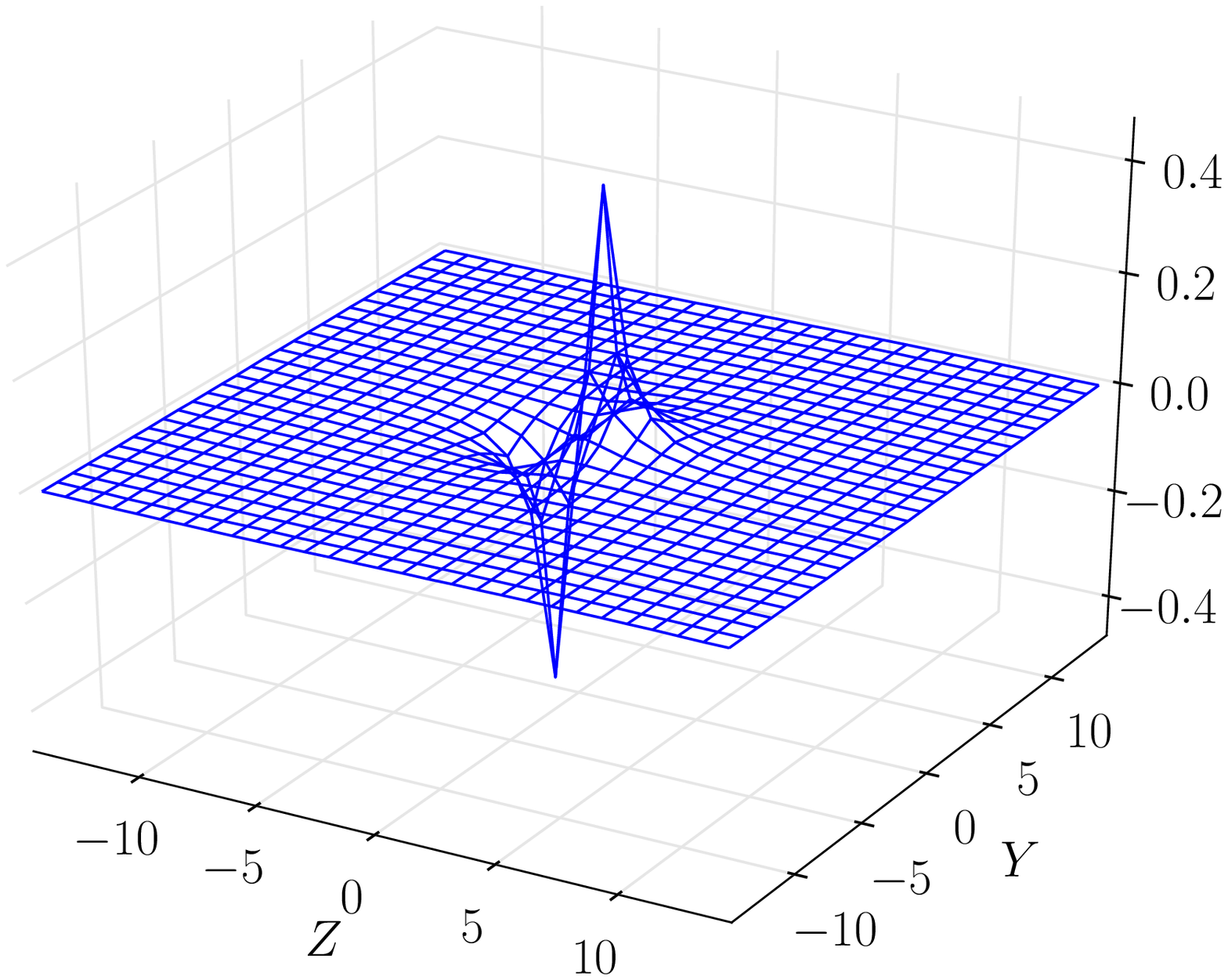}
    \includegraphics[width=.49\textwidth]{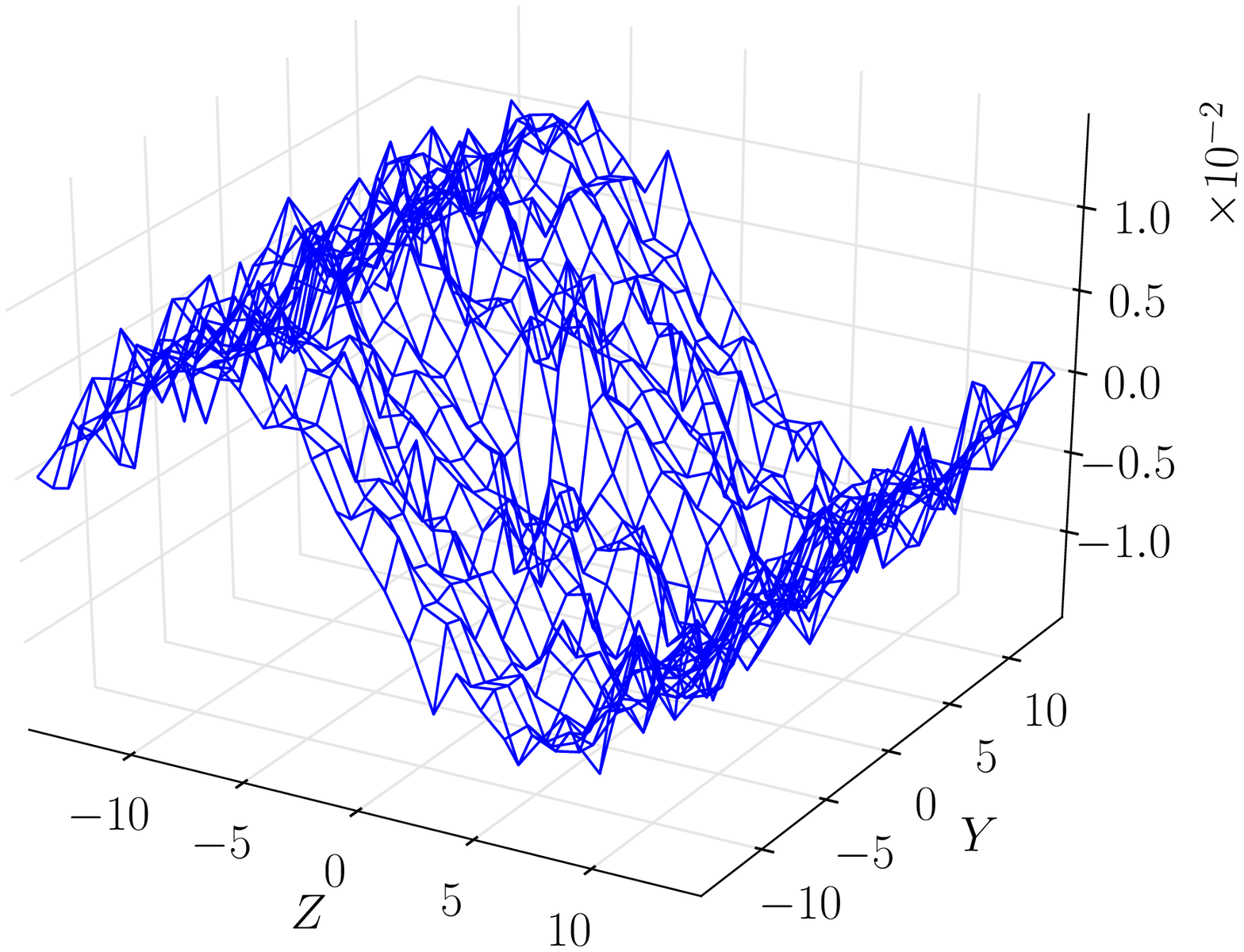}
    \caption{I=1 pion-pion wave functions on the ground (left) and first excited (right) states. Normalized to $1$. Computed at $t-t_0 = 10$ on the set $\kappa_{ud} = 0.1347$.}
    \label{fig:eigenstates}
\end{figure}

The ground state's wave function, Fig.~\ref{fig:eigenstates} (left), can be obtained using either source operator by saturation at large enough time separation. Taking $\pi\pi(\textbf p)$ as source operator, we see that the dominant contribution as time separation increases is from the rectangle diagram.

The first excited state's wave function is shown Fig.~\ref{fig:eigenstates} (right). We see that the dominant contribution is this time coming from the parallel diagram. The wave functions is obtained with a linear combination of the two source operators, which has for effect the cancellation of the peaked short-range contribution between the rectangle and triangle diagrams. However, while the signal from the ground state wave function is cancelled, the statistical noise remains and grows as $e^{\Delta E (t-t_0)}$ with $\Delta E$ the energy difference between the two lowest eigenstates.

An approximate potential is obtained by inverting the Schr\"odinger equation \eqref{eq:schrod_eq} with the BS wave functions computed on the lattice as input. The wave functions in Fig.~\ref{fig:eigenstates} unfortunately do not allow such a computation. The ground state wave function (left) is sharply peaked around the origin, leading to huge discretization errors when taking finite-difference Laplacian operator. The first excited state wave function (right) is extremely noisy due to large energy separation between the two lowest eigenstates and the noise is further enhanced by taking the Laplacian.

\begin{figure}
  \centering
  \includegraphics[width=.85\textwidth]{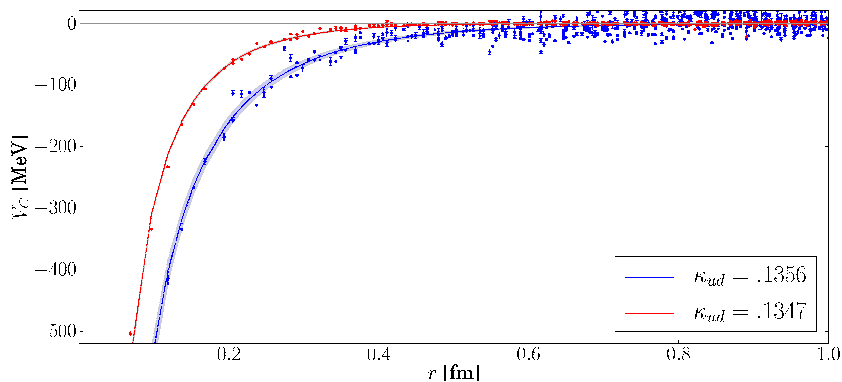}
  \caption{Central potential using only parallel diagrams. $t-t_0 = 12$. Fit by a Yukawa potential.}
  \label{fig:potential}
\end{figure}

Using the fact that the main contribution to the first excited state wave function is from the parallel diagram and that the other diagrams should only contribute to the short-range part of the potential, we show in Fig.~\ref{fig:potential} the effective central potential computed using only the parallel diagrams, on both sets of hopping parameters. We see that a simple Yukawa fit is in good agreement to the data even at surprisingly short range. The masses in the Yukawa fit are $1.53(9)$ GeV and $0.94(17)$ GeV, with corresponding rho masses of $1.37$ GeV and $1.10$ GeV, respectively.

\section{Summary and outlook}

We have shown preliminary results of the application of the potential method to the I=1 pion-pion system. The method, which has been successful in the study of baryon-baryon systems, encountered difficulties in this particular setup. First, the ground state being the rho meson, the wave function is very short-ranged and the computation of the potential leads to large discretization errors. Then, while the first excited state is a scattering state and likely to be well described by a potential, it is difficult to extract due to the large energy difference.

However, approaching the problem from a different perspective, the present results shed a new light on the qualitative understanding of the system. Furthermore, the above problems may be solved in the region where the rho meson is a resonance and not the ground state, since the scattering state will be simply extracted by saturation and the short-ranged component should become less important. In this case, the potential method could lead to competitive quantitative results. Further study at smaller pion masses will confirm or invalidate this expectation.

Numerical computations in this work were carried out on SR16000 at YITP in Kyoto University. We are also grateful for the authors and maintainers of CPS++ \cite{cps}, of which a modified version is used for measurement done in this work.


\begin{thebibliography}{99}
  \bibitem{pacscs-rho}{S. Aoki et al. [PACS-CS Collaboration], Phys. Rev. D 84, 094505 (2011) [arXiv:1106.5365].}
  \bibitem{lang-rho}{C. B. Lang et al., Phys. Rev. D 84, 054503 (2011) [arXiv:1105.5636].}
  \bibitem{bmw-rho}{J. Frison et. al. [BMW Collaboration], PoS LATT2010, 139 (2010) [arXiv:1011.3413].}
  \bibitem{feng-rho}{X. Feng, K. Jansen, and D. B. Renner, Phys. Rev. D 83, 094505 (2011) [arXiv:1011.5288].}
  \bibitem{luescher}{M. L\"uscher, Commun. Math. Phys. 105, 153 (1986); Nucl. Phys. B 364, 237 (1991).}
  \bibitem{rummukainen}{K. Rummukainen and S. A. Gottlieb, Nucl.Phys. B 450, 397 (1995) [hep-lat/9503028].}
  \bibitem{hal-proposal}{N. Ishii, S. Aoki and T. Hatsuda, Phys. Rev. Lett. 99, 022001 (2007) [arXiv:nucl-th/0611096].}
  \bibitem{hal-others}{T. Inoue et al. [HAL QCD Collaboration], Prog. Theor. Phys. 124, 591 (2010) [arXiv:1007.3559], \\
                       H. Nemura, N. Ishii, S. Aoki and T. Hatsuda, Phys. Lett. B 673, 136 (2009) [arXiv:0806.1094], \\
                       S. Aoki et al. [HAL QCD Collaboration], arXiv:0806.1094.}
  \bibitem{hal-theory}{S. Aoki, T. Hatsuda and N. Ishii, Prog. Theor. Phys. 123, 89 (2010) [arXiv:0909.5585].}
  \bibitem{configs}{T. Ishikawa et. al. [CP-PACS/JLQCD Coll.], Phys. Rev. D 78, 011502 (2008) [arXiv:0704.1937].}
  \bibitem{cps}{Columbia Physics System (CPS), http://qcdoc.phys.columbia.edu/cps.html}
\end{thebibliography}
\end{document}